\documentclass[a4paper,12pt]{article}
\usepackage{graphicx}
\usepackage{epsfig}

\begin{document}
\title{Momentum dependent higher partial wave interactions in Bose-Einstein condensate}
\author{Moumita Gupta\footnote{ E-mail:
 mou\_ mita\_ 81@yahoo.co.in} and Krishna Rai
Dastidar\footnote{E-mail: krishna\_raidastidar@yahoo.co.in}
\\  Department of Spectroscopy,
\\Indian Association for the Cultivation of Science
\\ Kolkata 700032, India}
\date{}
\maketitle

\section {Introduction}
For a very dilute Bose-Einstein condensate which is the case in
most of the experiments, the interatomic interaction is
sufficiently weak, the mean field Gross-Pitaevskii theory is a
logical tool to study such a system \cite {Dalfovo}. The physics
of the cold atom scattering is dominated by the two-body contact
interaction which is well described by the s-wave scattering
length. If we confine ourselves to the case of low momentum ($k$)
and the small values of scattering length ($a$) such that $ka<<1$,
the momentum dependence of the scattering amplitude of two
interacting atoms can be neglected \cite {Beliaev2, Ronen}.
However the value of the parameter $ka$ increases with the
increase in $a$, even for small momentum ($k$). Recent experiments
\cite {Cornish Rb85,Roati,Pollack} have explored the possibilities
of increasing the scattering length by exploiting magnetic
Feshbach resonance. In an experiment performed at JILA \cite
{Cornish Rb85} it was possible to confine $10^4$ atoms of
$^{85}Rb$ in a cylindrically symmetric trap at 100 $nK$. Around
the Feshbach resonance at a magnetic field B $\sim$ 155 Gauss, the
scattering length was varied from negative to very high positive
values $\sim$ 10000 $a_0$ by varying the magnetic field. In this
case the value of the parameter $ka$ becomes greater than $1$ at
and above $a=$ 3000 $a_0$. In this regime of $ka>1$, the momentum
dependence of scattering amplitude becomes important. The first
order theory of Beliaev \cite {Beliaev2} gives the momentum
dependence of scattering amplitude by applying field-theoretic
diagrammatic treatment to the zero-temperature homogeneous dilute
interacting Bose gas \cite {Beliaev2,Beliaev1}. The Beliaev theory
was generalized to give the temperature dependence of the
excitation spectrum of Bose gas \cite {Klyat, Shlyap1}. In the
regime $ka<<1$, the Beliaev's first order theory reduces to
Bogoliubov spectrum \cite{Bogo} for contact potential. The work of
Lee, Huang, Yang (LHY) \cite {Lee-Yang, Lee-Huang-Yang} on the low
temperature properties of dilute hard sphere gas considering the
perturbation theory gives the first correction to the Bogoliubov
mean field theory for dilute gases for relatively stronger
interaction. The energy correction obtained from Beliaev's second
order theory \cite {Beliaev2} coincides with the LHY result
assuming momentum independent scattering length. Brueckner and
Sadwa also determined the quantum depletion and correction to the
ground state energy of a homogeneous dilute Bose gas \cite
{Brueckner}. To study the ground state properties of a dilute Bose
gas with strong interactions momentum dependent scattering
amplitudes along with the contribution from LHY correction term is
important. For strongly interacting system ($ka>>1$), the higher
order partial waves likely to give significant contribution to the
scattering amplitudes. For $ka>1$, only the even higher partial
waves (e.g. d-wave ,g-wave etc.) will contribute to the atomic
interactions in the Bose gases. It has been mentioned by Beliaev
\cite {Beliaev2} that the d-wave contribution to energy spectrum
can be $\sim 10\%$. Here we will show that d-wave contribution to
the column density can be double of this value $\sim 20\%$ at the
centre of the trap for large values of scattering length.

In this letter we present the ground state energy functional of
trapped atomic Bose condensate including the effect of momentum
dependent scattering amplitude (with full generality) and also
including the LHY correction term. Higher partial wave
interactions have been incorporated in the energy functional. From
the ground state energy functional the time-independent equation
has been derived to give the condensate density. We have
considered $10^4$ $^{85}$Rb atoms which are cooled to a
temperature of 100 $nK$ with the scattering length varying from
3000 $a_0$ to 8700 $a_0$. The corresponding range of variation of
$ka$ is 1.15 to 3.33 and the values of peak gas-parameter
$x_{pk}\sim 10^{-2}$ ($x_{pk}$=$n(0)a^3$, where $n(0)$ is the peak
density of Bose gas and $a$ is the s-wave scattering length of
interatomic interaction) [see Table 1]. It is found that for
$ka>1$, the contribution of the momentum dependent scattering
(considering s, d and g waves) to the column density becomes
important and the nature of column density differs significantly
from that obtained for k-independent scattering (contact
interaction with LHY correction). The contribution from the g-wave
interaction to the column density is negligible (less than 1\%)
for ka= 3.33. As in the vicinity of Feshbach resonance the two and
three body loss rates play a crucial role \cite {Saito, Roberts,
Shlyap2, Braaten}, their effect have also been included
phenomenologically in the time-independent equations of the
condensate. The patterns of the column densities as observed
experimentally agree well with our results considering momentum
dependent scattering amplitude at B=157.2 Gauss ($a\approx 3000
a_0$).

\section {Theory}
Quantum scattering theory can be formulated in terms of partial
waves. The wave function for relative motion is written as
$\psi=e^{ikz}+f(\theta)e^{ikr}/r$, where $f(\theta)$ is the
scattering amplitude and k is the wave vector of the scattered
wave. The first term gives an incoming plane wave travelling along
z axis and the second term represents a radially outgoing
scattered wave \cite {Pethick-Smith}. Using partial wave expansion
the scattering amplitude for identical bosons can be expressed as
$$f(\theta)=(1/k)\sum_{l=0}^{\infty}(2l+1)exp(i\delta_l)sin\delta_l
P_l(cos\theta)\eqno(1)$$ Here, $l$= 0, 1, 2,... denotes the
contribution of s, p, d, . . . partial waves to the total
scattering amplitude, $\delta_l$ are the phase shifts associated
with each partial wave, and $P_l(cos\theta)$ are Legendre
polynomials. For identical bosons only even $l$ partial waves can
contribute to the scattering amplitude. According to Beliaev's
first order theory \cite {Beliaev2} the excitation spectrum of a
dilute Bose gas can be given by the dispersion relation
$$\epsilon^{(1)}_k=\sqrt{(\epsilon_k^0+2n_0f^{s}({k\over 2},{k\over 2})
-n_0f(0,0) {\hbar^2\over m})^2-(n_0f(k,0){\hbar^2\over
m})^2}\eqno(2)$$ where $\epsilon_k^0=\hbar^2k^2/(2m)$ is the
kinetic energy and $m$ is the atomic mass. Here only the forward
scattering of atoms are considered. The two distinct scattering
amplitudes (considering s-wave) are $f(0,0)= 4\pi a$ and $f(k,0)=
4\pi sin(ka)/k$. The symmetrized amplitude $f^{d}(k/2,k/2)$ for
higher partial wave (l=2) is added with $f^{s}(k/2,k/2)$ (for l=0)
to give the following dispersion relation
$$\epsilon^{(1)}_k=\sqrt{[\epsilon_k^0+2n_0\{(f^{s}({k\over 2},{k\over
2})+f^{d}({k\over 2},{k\over 2})\}-n_0f(0,0) {\hbar^2\over
m}]^2-(n_0f(k,0){\hbar^2\over m})^2}\eqno(3)$$ The values of
symmetrized scattering amplitudes considering s and d waves are
$f^{s}(k/2,k/2)= 4\pi (sin(ka)-i2sin^2(ka/2))/k$ and
$f^{d}(k/2,k/2)= -40\pi(cos\delta_2sin\delta_2+isin^2\delta_2)$;
where $\delta_2=tan^{-1}[\{3{ka\over 2}cos {ka\over
2}-(3-{k^2a^2\over4})sin{ka\over
2}\}/\{(3-{k^2a^2\over4})cos{ka\over 2}+3{ka\over 2}sin{ka\over
2}\}]$.

The LHY correction term for the ground state energy of a dilute
Bose gas \cite {Lee-Yang,Lee-Huang-Yang} is
$$E[n]={2\pi\hbar^2an\over m}{128\over 15}({na^3\over \pi})^{1/2}\eqno(4)$$
This term can also be obtained from second order Beliaev's theory
\cite{Beliaev2}. The ground state energy functional can be given
as
$$E[\psi]=\int d\vec r[-{\hbar^2\over
2m}\psi^*\nabla^2\psi+V_{tr}(\vec r)|\psi|^2+{1\over
2}\{g_1|\psi|^2+{(g_1^2-g_2^2)\over
2\epsilon_k^0}|\psi|^6+{256\over 15}{\hbar^2\over m}{\sqrt
\pi}a^{5/2}|\psi|^5\}\eqno(5)$$ where $g_1={\hbar^2\over
m}\{2(f^s(k/2,k/2)+f^d(k/2,k/2))-f(0,0)\}$ and $g_2={\hbar^2\over
m}f(k,0)$; $V_{tr}$ is the trapping potential. In $g_1$ and $g_2$
the imaginary part of the scattering amplitudes has not been taken
into account.

 By performing a functional variation with
respect to $\psi^*$ the Euler-Lagrange equation \cite{M2,M5} takes
the following form
$$[-{\hbar^2\over 2m}\nabla^2+V_{tr}+g_1|\psi|^2+{3\over 4}{g_1^2-g_2^2\over
\epsilon_k^0}|\psi|^4+{128\over 3}{\hbar^2\over m}{\sqrt
\pi}a^{5/2}|\psi|^3]\psi=\mu\psi\eqno(6)$$ where $\mu$ is the
chemical potential which accounts for the conservation of number
of particles. In Eq. (6) the two body and three body losses are
introduced phenomenologically to give
$$[-{\hbar^2\over 2m}\nabla^2+V_{tr}+g_1|\psi|^2+{3\over 4}{g_1^2-g_2^2\over
\epsilon_k^0}|\psi|^4+{128\over 3}{\hbar^2\over m}{\sqrt
\pi}a^{5/2}|\psi|^3-{i\hbar\over
2}(K_2|\psi|^2+K_3|\psi|^4)]\psi=\mu\psi\eqno(7)$$ where $K_2$ and
$K_3$ are the two body and three body recombination loss rate
coefficients, respectively.

At small momenta ($ka<1$), neglecting the d-wave scattering and
setting $f^{s}({k\over 2},{k\over 2})=f(k,0)=f(0,0)$ Eq. (6)
reduces to$$[-{\hbar^2\over
2m}\nabla^2+V_{tr}+g_1|\psi|^2+{128\over 3}{\hbar^2\over m}{\sqrt
\pi}a^{5/2}|\psi|^3]\psi=\mu\psi\eqno(8)$$ where
$g_1=4\pi\hbar^2a/m$. The Eq.(8) is known as modified
Gross-Pitaevskii equation for the atomic condensate \cite {M2}.

The trapping potential for the cylindrical trap with two angular
frequencies $\omega_\perp$ and $\omega_z$ is given as
$$V_{tr}({\bf r})={1\over 2}m(\omega_\perp^2r_\perp^2+\omega_z^2
z^2)\eqno(9)$$

The column density which is an accessible experimental quantity is
defined as $n_c(z)=\int {dr_\perp}|\psi(r_\perp,z)|^2$. A measure
of the extension of the condensate is the half width of the column
density $R_{1/2}$, defined as the z-value when
$n_c(z=R_{1/2})={1\over 2}n_c(z=0)$.

\section {Results and discussion}
Our aim here is to emphasize the significance of the momentum
(k)-dependent scattering amplitudes (both for s-wave and higher
partial wave scattering) to determine the ground state of the
system for large scattering lengths even at small values of
momentum. In order to do so we have considered a condensate of
$10^4$ $^{85}$Rb atoms confined in a cylindrical trap with radial
(axial) frequency $\omega_\perp$= 17.5 $Hz$ ($\omega_z$= 6.8 $Hz$)
with large gas parameter values ($\sim 10^{-2}$) as achieved in
the experiment of Cornish et al. \cite {Cornish Rb85}.
\begin{table}
\caption {Results for the ground-state properties of $10^4$
$^{85}$Rb atoms trapped in a cylindrically symmetric trap with
${\omega_\perp \over{2\pi}}$=17.5 Hz and $\omega_z$= 6.8 $Hz$.
Chemical potentials are in the units of $\hbar\omega_\perp$; half
widths are in the units of $\mu$m. Results are given for
k-dependent s-wave scattering [s(k)], k-dependent (s+d)-wave
scattering [s+d(k)] and k-independent s-wave scattering [s].}
\begin{tabular}{lllllllll}
\hline\hline
$a$ ($a_0$)&ka&& $\mu$ & $x_{pk}$  & half width\\
\hline
$3000$&1.15&s(k) & 12.47 & 3.89(-3)&21.01 \\
 && s+d(k) & 12.59 & 3.84(-3)&21.09 \\
 && s & 14.48 & 3.19(-3) & 22.42 \\\\
\hline
$5000$&1.92&s(k) & 12.74 & 1.49(-2)&22.52 \\
 && s+d(k) & 13.79 & 1.37(-2)&23.09 \\
 && s & 18.77 & 9.72(-3) & 25.88 \\\\
\hline
$7000$&2.68&s(k) & 11.51 & 3.61(-2)&23.72 \\
 && s+d(k) & 15.15 & 2.97(-2)&25.15 \\
 && s & 22.56 & 1.96(-2) &28.62 \\\\
\hline
$8700$&3.33&s(k) &11.16& 5.89(-2)&25.05 \\
 && [s+d](k) & 17.79 & 4.45(-2)&27.32 \\
 && s & 25.57 & 3.08(-2) & 30.58 \\\\
 \hline\hline
\end{tabular}
\end{table} We have made an attempt to explain some
experimental results in light of the k-dependent scattering
phenomena. In Table I we list the values of the chemical potential
($\mu$), peak gas-parameter ($x_{pk}$) and also the half widths of
the column density distributions (to be demonstrated in Fig. 1)
along with the values of the parameter $ka$ considering
k-independent s-wave and also k-dependent s- and (s+d)- wave
scattering. For the range of $a$ considered here the parameter
$ka$ is greater than 1 and assumes the maximum value of 3.33 for
$a$= 8700 $a_0$. As expected the differences between k-dependent s
and (s+d) results increases with the increase in $ka$ and are of
the order of  59\% for $\mu$, 24\% for $x_{pk}$ and 9\% for the
half width at $a=8700 a_0$. The differences between these results
(considering momentum-dependent scattering) and those for
k-independent s-wave scattering are also enhanced at large values
of $ka$. Note that for the k-dependent s-wave scattering $\mu$
decreases with the increase in $a$ after 5000 $a_0$. This effect
of the negative contribution to the chemical potential is
nullified due to the inclusion of d-wave in the k-dependent
scattering amplitude.

The column densities considering k-dependent s- and (s+d)-wave
[obtained by solving Eq. (6)] and k-independent s-wave [obtained
by solving Eq. (8)] atom-atom scattering are presented in Fig.
1(a), (b) and (c) for $a$= 3000 $a_0$, 7000 $a_0$ and 8700 $a_0$
respectively. The dotted and solid lines correspond to the
k-dependent s-wave and (s+d)-wave scattering results; circles are
the k-independent s-wave results; the red dashed lines give the
column densities with (s+d)-wave scattering considering two-body
($K_2$) and three-body ($K_3$) losses [obtained by solving Eq.
(7)]. The values of $K_2$ and $K_3$ are taken from the Fig. 2 in
Ref. \cite {Roberts} where two- and three- body loss rates of
$^{85}$Rb atoms are shown as a function of magnetic field near
Feshbach resonance. Column densities considering k-dependent s-
and (s+d)-wave scattering almost coincides at $a= 3000 a_0$ ($ka=
1.15$), whereas for larger values of $ka$, the (s+d)-wave
scattering lowers the central column density with an expansion in
the density distribution leading to a larger half width (as shown
in Table 1) than the column density corresponding to k-dependent
s-wave scattering. At $a=8700$ $a_0$ the deviation is 20.6\% at
z=0, which can be experimentally detected where the accuracy of
measurement is higher \cite {Stampern-Kurn}. The column density
considering k-independent s-wave scattering is consistently
lowered in the centre than the k-dependent results and the half
width of the atomic cloud condensate is larger.

In Fig. 2 we have compared our results for $a=$ 3000 $a_0$ with
that obtained in the experiment \cite {Cornish Rb85} for B= 157.2
Gauss. In the conditions of this experiment the scattering length
corresponding to B=157.2 Gauss is $\sim 3000 a_0$ (as obtained by
manual interpolation from Fig. 1 of Ref. \cite {Cornish Rb85}). We
have solved Eqs. (6) and (7) by taking $a= 3000 a_0$. In Fig. 2
the circles are the plots of the experimental data obtained by the
interpolation of the condensate column density curve taken from
Fig. 3(d) in Ref. \cite {Cornish Rb85}. The solid line indicates
column densities for k-dependent (s+d)- wave scattering
considering two and three body losses [obtained by solving Eq.
(7)] and the dotted line gives the same without considering any
losses [obtained by solving Eq. (6)]. The theoretical results are
fitted with the experimental data at the second point from the z=0
axis (at z=4.03 $\mu m$) and the deviation between experimental
and theoretical results at z=0 is about 10\%. By comparison we
find that our theoretical results are in good agreement with the
experimental data up to z= 27 $\mu m$, but the long tail obtained
in the experiment beyond 27 $\mu m$ could not be reproduced. As
the half width of the of the column density for k-independent
s-wave interaction is larger than that for the k-dependent
interaction [see Table 1 and Fig. 1(a)], it is obvious that the
column density corresponding to the k-independent scattering will
deviate outwardly from the experimental data. We have repeated the
calculation with $a= 3100 a_0$ and $2900 a_0$ (as the error in the
interpolation of the curve in Fig. 1 of Ref. \cite {Cornish Rb85}
is within 90 $a_0$) and found the results are almost coincident
with those for 3000 $a_0$. Increases in the half widths obtained
from our theoretical column densities due to the increase in $a$
from 3000 $a_0$ to 8700 $a_0$ are 29.5\% and 36.4\% for
k-dependent (s+d)-wave and k-independent s-wave scattering [see
Table 1]. In the experiment $a$= 3000 $a_0$ and 8700 $a_0$
correspond to B= 157.2 Gauss and 156.4 Gauss [obtained by manual
interpolation of Fig.1 in Ref. \cite{Cornish Rb85}]. The increase
in the half width of the column density is $\sim 15\%$ due to the
change in B from 157.2 G to 156.4 G. Smaller increase in the half
widths due to increase in $a$ in the case of k-dependent results
than the k-independent results indicates that k-independent
results are closer to the experiment. In this study we have not
considered the effect of coherent loss process and also the finite
temperature corrections which may affect the column density for
the large values $a$.

To summarize, we have shown that the inclusion of the momentum
dependent scattering amplitudes due to s-wave and higher partial
waves are important to determine the ground state properties of
the condensate when $ka>1$ and $x_{pk}\sim 10^{-2}$.
Experimentally detectable significant quantitative differences are
found between the results considering momentum independent and
momentum dependent scattering. Theoretical column densities for
$^{85}$Rb atom considering (s+d)-wave scattering are found to
exhibit good agreement with the experimental results for $a=$ 3000
$a_0$.

{\bf Figure captions}
\begin{description}
\item Fig. 1 Column densities of $10^4$ $^{85}$Rb atoms at $a$ =
3000 $a_0$ (a),  $a$ = 7000 $a_0$ (b) and $a$ = 8700 $a_0$ (c) as
a function of axial distance for a cylindrical trap with
$\omega_{\perp}/2\pi$ = 17.5 Hz and $\omega_z/2\pi$ = 6.8 Hz. The
vertical axis is multiplied by $2\times
a_{\perp}^2/N$($a_{\perp}=\sqrt{\hbar/(m\omega_\perp)}$). The
dotted and solid lines represent the densities with k-dependent
s-wave and (s+d)-wave scattering; circles are the k-independent
s-wave results; the red dashed lines give the column densities
considering two-body ($K_2$) and three-body ($K_3$) losses.

\item Fig. 2  Column densities of $10^4$ $^{85}$Rb atoms at $a$ =
3000 $a_0$ (corresponding to B=157.2 gauss in Ref. \cite{Cornish
Rb85}) as a function of axial distance for a cylindrical trap with
$\omega_{\perp}/2\pi$ = 17.5 Hz and $\omega_z/2\pi$ = 6.8 Hz. The
solid line gives the result for k-dependent (s+d)-wave scattering
considering two and three body losses and the dotted line gives
the column density without considering any losses. Circles are the
plots of the experimental condensate column density taken from
Fig. 3(d) in Ref. \cite {Cornish Rb85}.

\end{description}


\begin{thebibliography}{widest-label}
\bibitem {Dalfovo} F. Dalfovo, S. Giorgini, L. Pitaevskii, and S. Stringari, Rev.
Mod. Phys. {\bf 71}, 463 (1999)
\bibitem {Beliaev2} S. T. Beliaev, Zh. Eksp. Teor. Fiz. {\bf 34}, 433 [Sov. Phys. JETP {\bf 7}, 299 (1958)]
\bibitem {Ronen} S. Ronen, J. Phys. B: At. Mol. Opt. Phys. {\bf 42}, 055301 (2009)
\bibitem {Cornish Rb85} S. L. Cornish, N. R. Claussen, J. L. Roberts, E. A. Cornell and
C. E. Wieman, Phys. Rev. Lett. {\bf 85}, 1795 (2000)
\bibitem {Roati} G. Roati, M. Zaccanti, C. D'Errico, J. Catani, M. Modugno, A.
Simoni, M. Inguscio, and G. Modugno, Phys. Rev. Lett. {\bf 99},
010403 (2007)
\bibitem {Pollack} S. E. Pollack, D. Dries, M. Junker, Y. P. Chen, T. A. Corcovilos,
and R. G. Hulet, Phys. Rev. Lett. {\bf 102}, 090402 (2009).
\bibitem {Beliaev1} S. T. Beliaev, Zh. Eksp. Teor. Fiz. {\bf 34},
417 (1958) [Sov. Phys. JETP {\bf 7}, 289 (1958)]
\bibitem {Klyat} V. I. Klyatskin, Russian Phys. Journal {\bf 9}, 100 (1966)
\bibitem {Shlyap1} P. O. Fedichev and G. V. Shlyapnikov, Phys. Rev. Lett. {\bf 58}, 3146 (1998)
\bibitem {Bogo} N. N. Bogoliubov, J. Phys. (Moscow) {\bf11}, 23 (1947)
\bibitem {Lee-Yang} T. D. Lee and C. N. Yang, Phys. Rev.
{\bf105}, 1119 (1957)
\bibitem {Lee-Huang-Yang} T. D. Lee, K. Huang and C. N. Yang, Phys. Rev.
{\bf106}, 1135 (1957)
\bibitem {Brueckner} K. A. Brueckner and K. Sawada, Phys. Rev. {\bf 106}, 1117 (1957)
\bibitem {Saito} H. Saito and M. Ueda, Phys. Rev. A {\bf 65}, 033624 (2002)
\bibitem {Roberts} J. L. Roberts et al., Phys. Rev. Lett. {\bf 85}, 728 (2000)
\bibitem {Shlyap2} P. O. Fedichev, M. W. reynolds and G. V. Shlyapnikov, Phys. Rev. Lett. {\bf 77}, 2921 (1996)
\bibitem {Braaten} E. Braaten and H. W. Hammer, Phys. Rev. Lett. {\bf 87}, 160407 (2001)
\bibitem {Pethick-Smith} C. J. Pethick and H. Smith, Bose-Einstein Condensation in Dilute
Gases, Cambridge University Press (2002)
\bibitem {M2} M. Gupta and K. R. Dastidar, J. Phys. B: At. Mol. Opt. Phys. {\bf41}, 195302 (2008)
\bibitem {M5} M. Gupta and K. R. Dastidar, Phys. Rev. A {\bf 81}, 063631 (2010)
\bibitem {Stampern-Kurn} D-M Stampern-Kurn et al., Phys. Rev. Lett. {\bf 81}, 500 (1998)
\end{thebibliography}
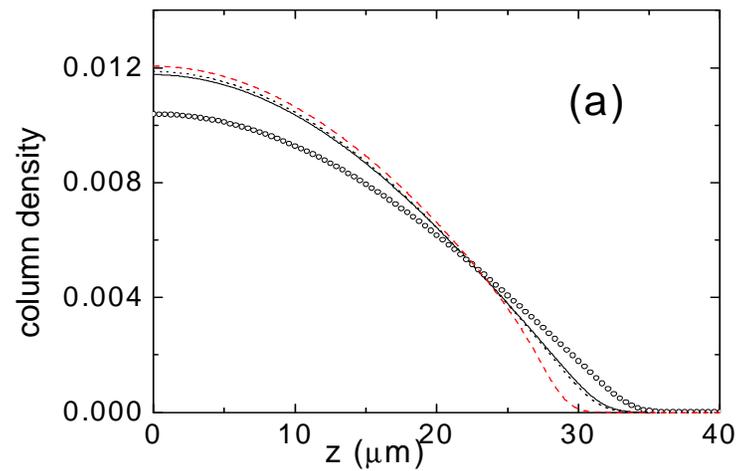
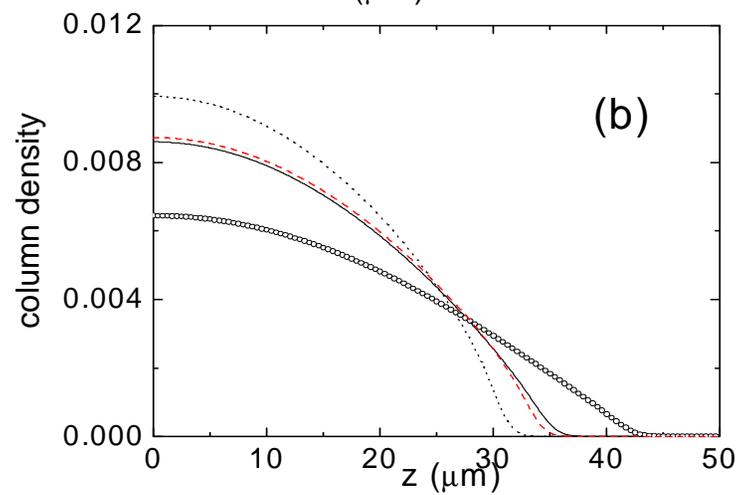
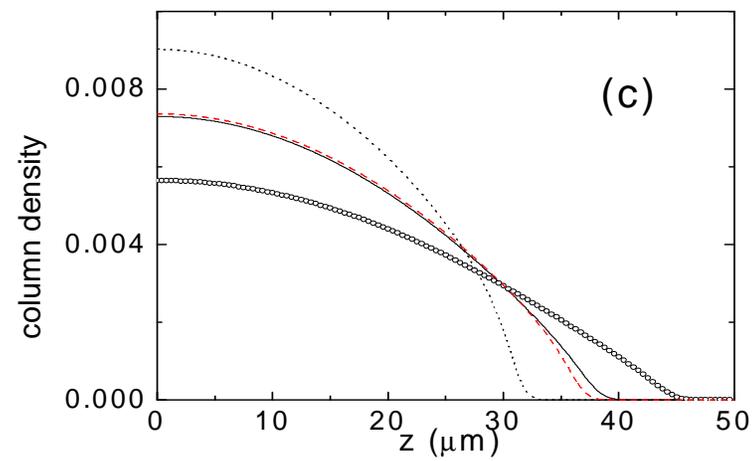
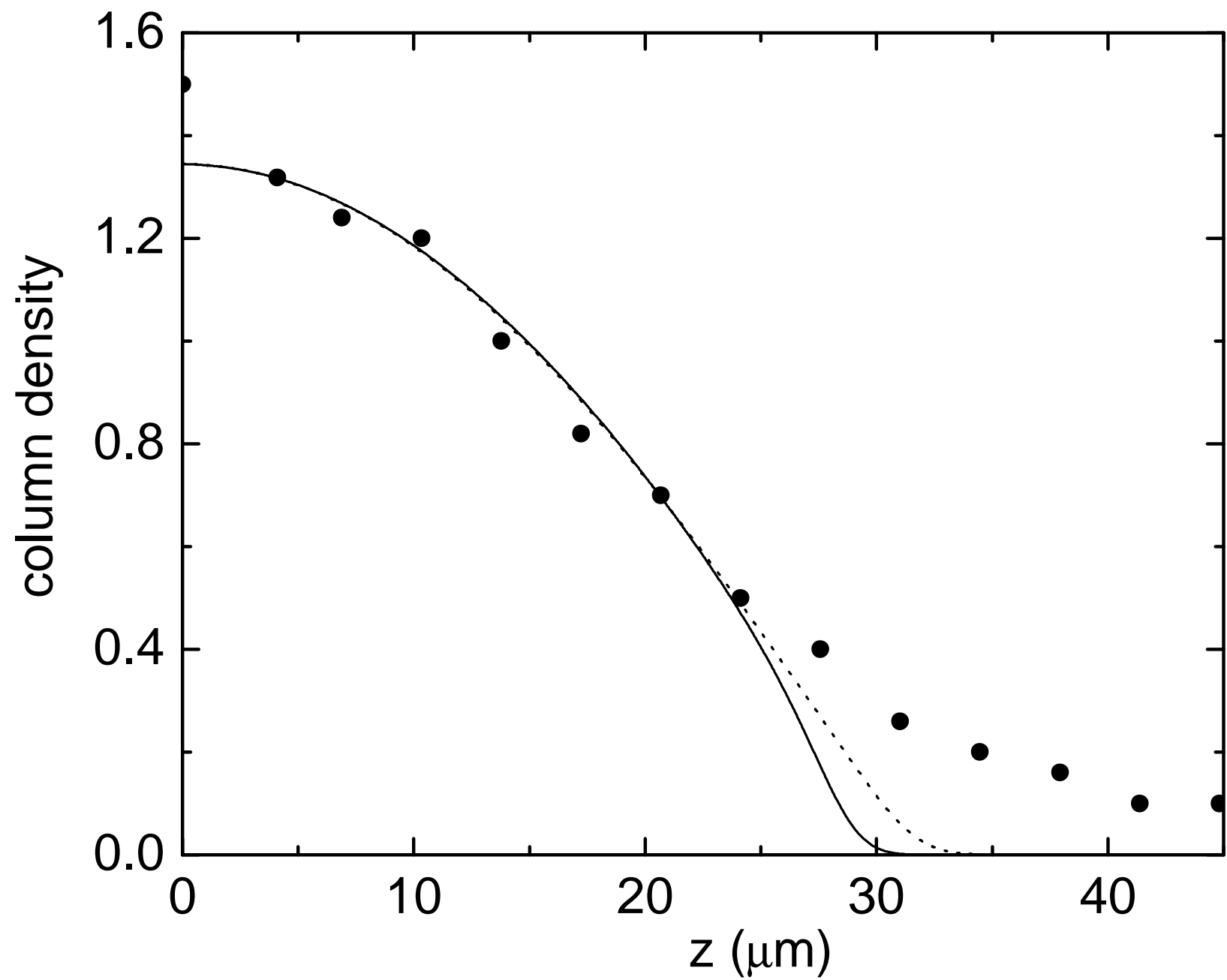
\end{document}